\documentclass[sigconf]{acmart}

\usepackage{subcaption}
\usepackage{tabularx}

\newcolumntype{Y}{>{\centering\arraybackslash}X}

\AtBeginDocument{%
  \providecommand\BibTeX{{%
    \normalfont B\kern-0.5em{\scshape i\kern-0.25em b}\kern-0.8em\TeX}}}

\copyrightyear{2022} 
\acmYear{2022} 
\setcopyright{acmcopyright}
\acmConference[DAC '22]{Proceedings of the 59th ACM/IEEE Design Automation Conference (DAC)}{July 10--14, 2022}{San Francisco, CA, USA}
\acmBooktitle{Proceedings of the 59th ACM/IEEE Design Automation Conference (DAC) (DAC '22), July 10--14, 2022, San Francisco, CA, USA}
\acmPrice{15.00}
\acmDOI{10.1145/3489517.3530504}
\acmISBN{978-1-4503-9142-9/22/07}

\begin{document}

\title{SALO: An Efficient Spatial Accelerator Enabling Hybrid Sparse Attention Mechanisms for Long Sequences}

\author{Guan Shen, Jieru Zhao\footnotemark[1], Quan Chen, Jingwen Leng, Chao Li, Minyi Guo}
\authornote{Jieru Zhao and Minyi Guo are corresponding authors.}
\affiliation{
\institution{Department of Computer Science and Engineering, Shanghai Jiao Tong University}
\city{}
\country{}
}
\email{{shenguan, zhao-jieru}@sjtu.edu.cn, {chen-quan, leng-jw, lichao, guo-my}@cs.sjtu.edu.cn}

\begin{abstract}
  The attention mechanisms of transformers effectively extract pertinent information from the input sequence. However, the quadratic complexity of self-attention w.r.t the sequence length incurs heavy computational and memory burdens, especially for tasks with long sequences. Existing accelerators face performance degradation in these tasks. To this end, we propose SALO to enable hybrid sparse attention mechanisms for long sequences. SALO contains a data scheduler to map hybrid sparse attention patterns onto hardware and a spatial accelerator to perform the efficient attention computation. We show that SALO achieves $17.66$x and $89.33$x speedup on average compared to GPU and CPU implementations, respectively, on typical workloads, i.e., Longformer and ViL.
\end{abstract}

\keywords{}
\settopmatter{printacmref=false}

\maketitle
\pagestyle{plain}
\pagenumbering{gobble}
\section{Introduction}
Nowadays, models based on Transformer \cite{vaswani2017attention} have achieved an extremely high performance in deep learning research. In the area of natural language processing, Transformer and its variants, like BERT \cite{devlin2018bert} and GPT-3 \cite{brown2020language}, have outperformed other models based on RNN and CNN in most tasks. Inspired by the great ability that transformers have shown in NLP tasks, researchers have started to migrate the network to other fields like computer vision and recommender system recently. Vision Transformer (ViT) \cite{dosovitskiy2020image} is one of the representative works which apply the transformer directly to sequences of image patches.

Transformer and its variants achieve state-of-the-art results by capturing contextual information from the entire sequence using the self-attention mechanism. Each self-attention block takes three matrices, namely $Q$ (query), $K$ (key), and $V$ (value), as its inputs. These three matrices will be used to capture the relation between tokens in a given sequence. The computation of attention includes two matrix multiplications, a scaling operation and a softmax operation. The two matrix multiplications introduce the quadratic computational complexity with respect to the sequence length. As transformers are being extensively applied in many areas, the model size increases greatly, together with the lengths of input sequences. 
For example, BERT-base \cite{devlin2018bert} and GPT-3 \cite{brown2020language} have 110 million and 175 billion parameters, respectively. Due to the quadratic complexity of self-attention w.r.t the sequence length, the model complexity is further exacerbated in scenarios with long input sequences. Such considerable workload brings about heavy computational and memory burdens, making it difficult to train and deploy these models, especially for tasks with long input sequences.

To remedy this issue, several models with hybrid sparse attention mechanisms are proposed to handle long sequences, such as Longformer \cite{beltagy2020longformer} (up to 16384 tokens in a sequence) and ViL \cite{zhang2021multi} (up to $96\times96$ patches in an image). They hybridize two variants of the attention mechanism and perform the computation of a local window attention and a task-motivated global attention. Such a hybrid sparse attention mechanism successfully reduce the complexity of attention to a linear level. This linear complexity significantly alleviates the memory burden, making it possible to train models with long sequences. However, since the hybrid sparse attention mechanism is not directly supported by the highly optimized GEMM kernels on CPU and GPU, the inference speed is limited, requiring efficient hardware acceleration to achieve a higher speed.

Existing attention accelerators \cite{li2020ftrans, wang2021spatten, ham20203, lu2021sanger} meet performance bottlenecks given long input sequences. FTRANS \cite{li2020ftrans} compresses weights of transformers while computing the full attention without sparsity. SpAtten \cite{wang2021spatten} applies coarse-grained pruning to remove tokens and heads with a relatively low pruning ratio, which cannot effectively shorten the sequence length. $A^3$ \cite{ham20203} and Sanger \cite{lu2021sanger} accelerate the attention by utilizing dynamic sparse patterns, which can incur large additional memory/computation overhead in the case of long sequences. 
To this end, we propose SALO to efficiently accelerate the attention in transformers for tasks with long input sequences. The main contributions are summarized as follows:
\begin{itemize}
    \item We survey popular sparse attention mechanisms and explore the common computation patterns they share. Moreover, we analyze their potential of data reuse during computation.
    \item We propose SALO, an efficient spatial accelerator based on systolic array, to enable hybrid sparse attention mechanisms for long sequences. We utilize SALO to fully exploit the data reuse in the hybrid sparse attention mechanisms.
    \item We propose a data scheduler to transform a given sparse attention pattern into the hybrid sparse attention mechanism supported by SALO, through data splitting and data reordering. This ensures the flexibility of SALO.
\end{itemize}
Experimental results show that SALO achieves $17.66$x and $89.33$x speedup on average compared to GPU and CPU implementations, respectively, on typical workloads like Longformer and ViL.
\vspace{-0.1cm}
\section {Background and Motivation}

\subsection{Attention and Performance Analysis}
\begin{figure}
\vspace{-0.2cm}
  \centering
    \includegraphics[width=0.4\textwidth]{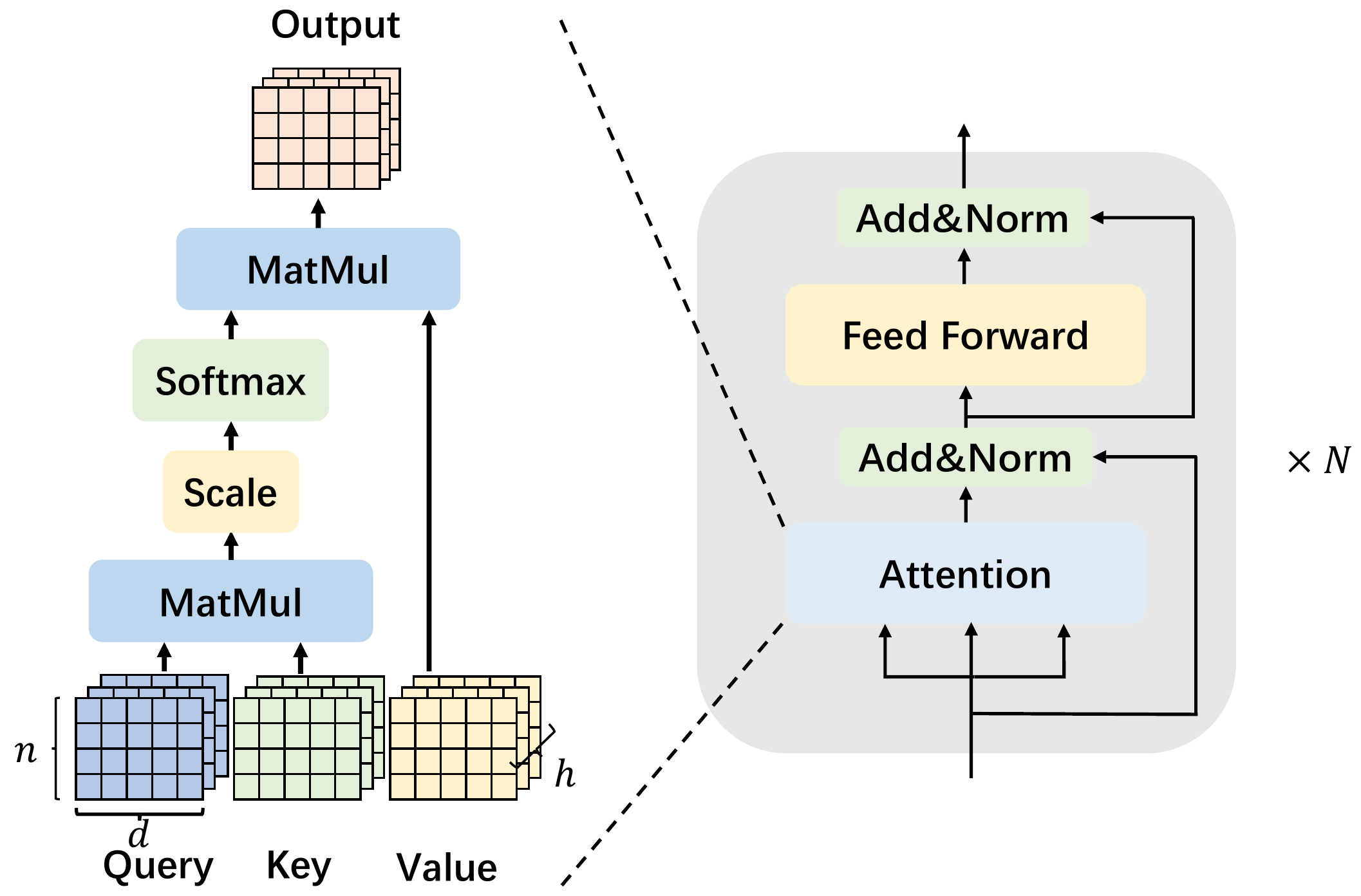}
    \caption{The computation of vanilla attention mechanism}
    \label{fig:attention}
    \vspace{-0.3cm}
\end{figure}

The attention mechanism contributes a lot to the superior performance of Transformers. 
Figure \ref{fig:attention} shows the detailed computation stages of the vanilla attention mechanism. Taking a sequence consisting of $n$ tokens as input, the representation of each token will be projected to three $d$-dimension vectors. And these $d$-dimension vectors of $n$ input tokens form three $n\times d$ matrices, namely $Q$ (query), $K$ (key), and $V$ (value), which are the inputs of the attention module. There are $h$ sets of inputs that stand for $h$ heads in multi-head attention.
The process of computing attention for each head contains several steps. First, $Q$ and $K$ are multiplied to get a $n\times n$ matrix $S$. This procedure can also be described at the vector scale: the dot product of each vector $q_i$ and each vector $k_j$ stands for an element in matrix $S$, measuring the impact that token $j$ has on token $i$. Second, the resultant matrix $S$ is scaled by $\sqrt{d}$ and for each row in $S$, softmax is then applied to get a normalized attention distribution, represented by a $n \times n$ matrix $S'$. The detailed calculation of softmax is shown below. 
\begin{equation}
\label{equ:softmax}
S'_{ij} = \mathrm{softmax}(S)_{ij}= \frac{\mathrm{exp}(S_{ij})}{\sum_k \mathrm{exp}(S_{ik})}
\end{equation}
Finally, the attention matrix $S'$ and the $V$ matrix are multiplied to obtain the output result. At the vector scale, the output of each token is the weighted sum of the value vectors according to the attention distribution. The above procedure is one "head" in multi-head attention. The output vectors from all heads will then be concatenated to get the final output.

We could see that the two matrix multiplications dominate the computational complexity of the attention mechanism: the first multiplies $Q (n\times d)$ and $K^T (d\times n)$ with complexity $O(n^2\cdot d)$, and the second multiplies $S' (n\times n)$ and $V (n\times d)$ with complexity $O(n^2\cdot d)$. In other words, the workload of the attention mechanism is quadratic with respect to $n$, i.e., the length of sequence. We test the latencies of a single attention layer of a pretrained BERT-base with different sequence length $n$, on a GTX-1080Ti. As the sequence length increases, the latency grows rapidly. The latency with $n=8192$ is 145.70ms, which is approximately $16$ times higher compared to a 9.20ms latency with $n=2048$, which shows the quadratic growth. Such a heavy workload makes it challenging to accelerate the attention mechanism for long sequences.

\subsection{Existing Accelerators}
Several existing accelerators \cite{li2020ftrans, ham20203, wang2021spatten, lu2021sanger} have been proposed to accelerate the attention computation. FTRANS \cite{li2020ftrans} compresses its model with the block-circulant matrix-based weight representation, but brings no sparsity to the quadratic attention computation. $A^3$ \cite{ham20203} accelerates the attention mechanism by dynamically selecting the elements in the key matrix that are more likely to influence the dot product of $q$ and $k$. However, this method stores the whole preprocessed key matrix on the SRAM buffer, making it difficult to scale up when the memory storage is insufficient given long input sequences. SpAtten \cite{wang2021spatten} adopts coarse-grained pruning techniques to prune tokens and heads. However, its relatively low pruning ratio leads to low sparsity and cannot effectively reduce the input size when the sequence length is long. Sanger \cite{lu2021sanger} approximates the attention mechanism by masking out elements below the threshold in the predicted score matrix $\hat{S}$. The pre-computation of the mask will be expensive in the case of long sequences due to the quadratic attention matrix multiplications. Moreover, these methods introduce intrusive modification of the attention mechanism to leverage the sparsity. Users need to tune task-specific parameters for a suitable trade-off between accuracy and speed. Our work, in contrast, can accelerate the attention computation with long input sequences efficiently and fit in various sparse attention mechanisms without any modification of the algorithm.

\subsection{Sparse Attention Mechanisms}
\label{subsec:sparse_attention_mechanisms}
\begin{figure}
\vspace{-0.3cm}
  \centering
  \begin{subfigure}[b]{0.13\textwidth}
    \includegraphics[width=\textwidth]{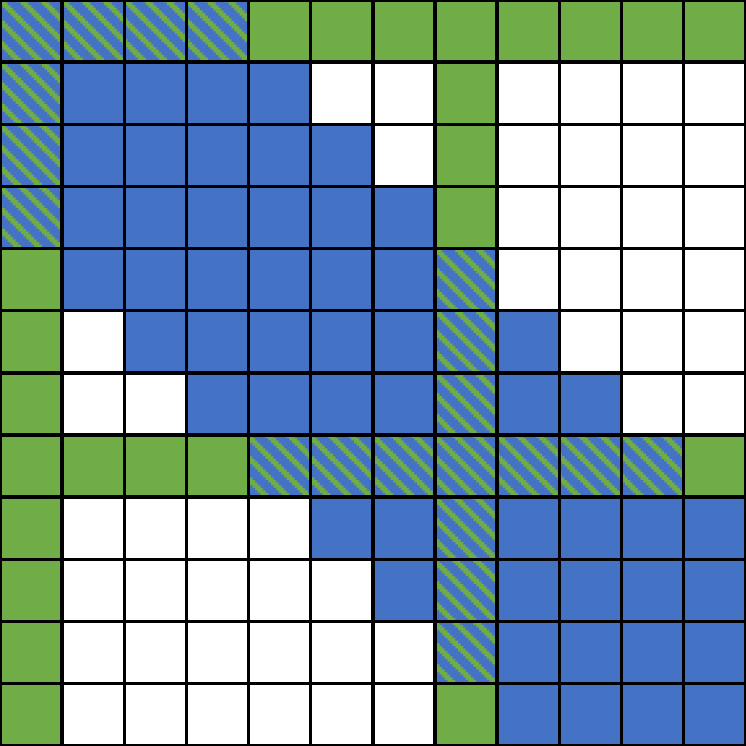}
    \caption{}
    \label{fig:longformer}
  \end{subfigure}
  \begin{subfigure}[b]{0.1653\textwidth}
    \includegraphics[width=\textwidth, page=4]{figure/pattern_v3.pdf}
    \caption{}
    \label{fig:star_transformer}
  \end{subfigure}
  \quad
  \begin{subfigure}[b]{0.13\textwidth}
    \includegraphics[width=\textwidth]{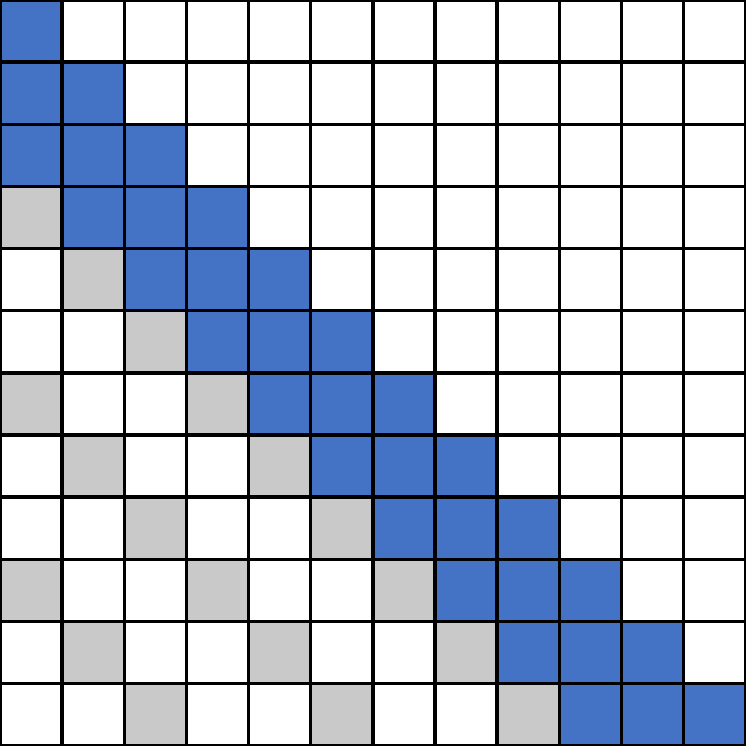}
    \caption{}
    \label{fig:sparse_transformer}
  \end{subfigure}
  \caption{Sparse Attention Mechanisms (a) Longformer (b) Star Transformer (c) Sparse Transformer }
  \label{fig:sparse}
  \vspace{-0.5cm}

\end{figure}

To remedy the full attention's quadratic dependency on the sequence length, several models with sparse attention mechanisms \cite{beltagy2020longformer, zhang2021multi, guo2019star, child2019generating} have been proposed. They successfully reduce the complexity of the attention module to linear or sub-quadratic one, enabling longer sequences for training and inference. Moreover, these variants introduce locality into Transformer, which is a good inductive bias for data like long texts and images. We list some of the representative sparse attention mechanisms in Figure \ref{fig:sparse} (We flatten the 2D patches of an image into a 1D sequence in Figure \ref{fig:sparse_transformer}). These patterns specify pairs of $q$ and $k$ vectors attending to one another, and can be regarded as masks on the resultant matrix $S$: if a position $(i,j)$ in the pattern is blank, the element $S_{ij}$ at corresponding position in $S$ will be ignored in the following calculation. In other words, the impact that token $j$ has on token $i$ is ignored. For example, in Figure \ref{fig:star_transformer}, the 6th row contains 3 blue blocks from the 5th column to the 7th column, which indicates that the query $q_6$ should attend to $k_5$, $k_6$, $k_7$. We could see that these sparse attention mechanisms share some sparse patterns in common and we mark them with different colors. Our insight is that it is possible to fully exploit the data reuse in these common sparse patterns, which motivate us to design our accelerator enabling hybrid sparse attention mechanisms.

\noindent\textbf{Sliding Window Attention.}
As the most common pattern in sparse attention mechanisms, sliding window attention (highlighted in blue) is an intuitive solution emphasizing the importance of the local context. Given a relative position range $[a, b]$, we could generate a sliding window attention with a fixed window size $w=b-a+1$, in which each query $q_i$ attends to the keys $k_j$ satisfying $a \leq j-i \leq b$.
We discover that, for the adjacent queries $q_i$ and $q_{i+1}$, there is only one different key vector between the sets of the key vectors they attend to, as highlighted in yellow and red positions in Fig. \ref{fig:star_transformer}.

\noindent\textbf{Dilated Window Attention.}
Similar to the dilated convolution in CNN, dilated window attention (highlighted in gray in Fig. \ref{fig:sparse_transformer}) is an extension to sliding window attention with a new parameter, dilation $d$, as the size of the gap in the sliding window. The dilated window attention can extend the receptive fields and represent the attention along y-axis in 2D scenarios (like image). The reuse of key vectors exists between queries $q_i$ and $q_{i+d}$.

\noindent\textbf{Global Attention.} 
Global attention plays a crucial role in capturing the global information by computing the full attention with some pre-selected special tokens. The choices of the global tokens are task-specific. The queries of the global tokens attend to all the keys in the sequence and the keys of the global tokens attend to all the queries in the sequence, which implies data reuse along the whole sequence.

\section{Framework Overview}
\begin{figure}[h]
    \vspace{-0.1cm}
  \centering
    \includegraphics[width=0.48\textwidth]{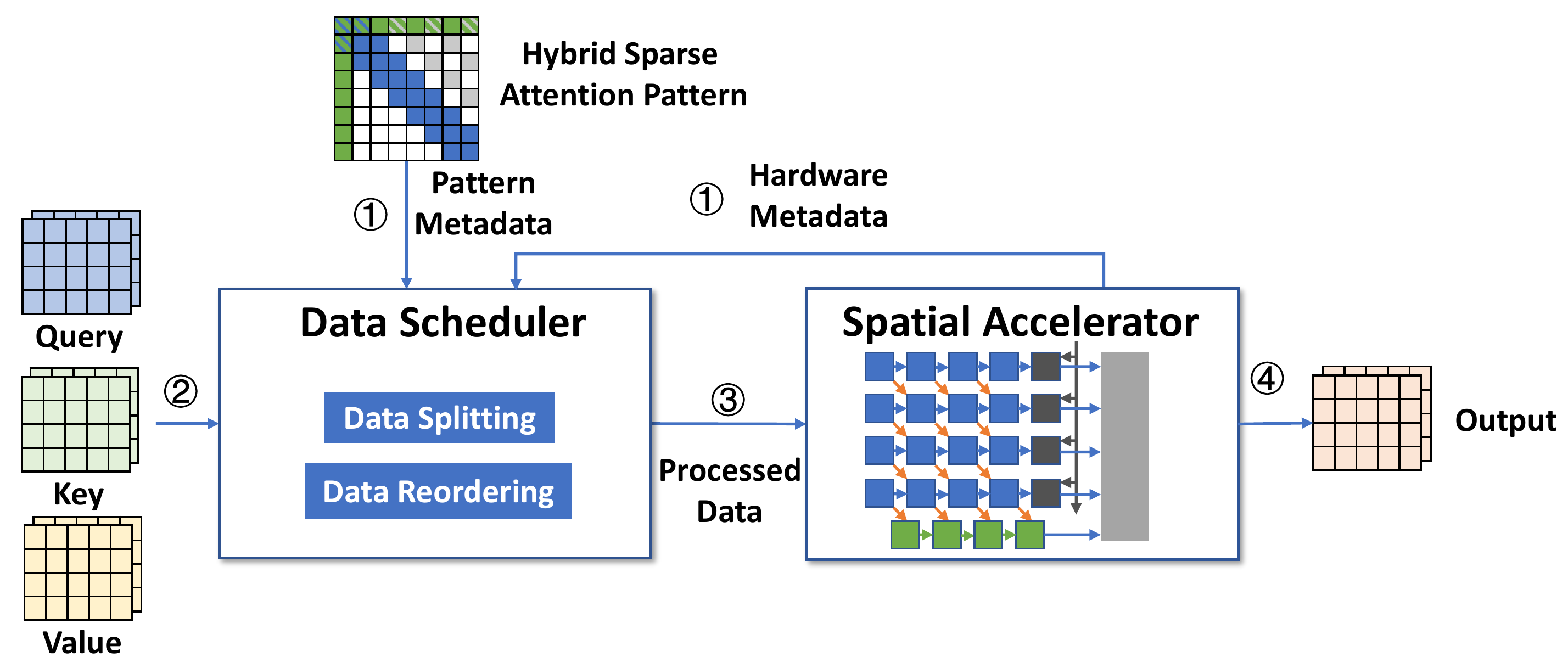}
    \caption{The framework overview of SALO}
    \label{fig:overview}
    \vspace{-0.2cm}
\end{figure}
Figure \ref{fig:overview} presents the overview of SALO, which consists of a data scheduler and a spatial accelerator. The data scheduler receives the metadata about the hybrid sparse attention pattern and the spatial accelerator. The metadata of the hybrid sparse attention pattern can describe the attention pattern, such as the window size $w$ of sliding window attention and the dilation $d$ of dilated window attention. The metadata of the spatial accelerator consists of the size of the PE array, and the number of global PE rows and columns. When the input (Query, Key, and Value) of the attention block in the model arrives, the data will be split to fit in the size of the spatial accelerator. As for dilated window attention, an extra reordering step will be involved. The details about the data scheduler will be discussed in Section \ref{subsec:data_scheduler}. The spatial accelerator contains a PE array (blue), a global PE row (green), a global PE column (black) and peripheral circuits (grey) in the figure. The PE array is the main part to compute the sliding window attention with specially designed diagonal connections to maximize the data reuse. The global attention part in the hybrid sparse attention mechanism is computed with the global PE column and the global PE row. 
The detailed design of the accelerator will be presented in Section \ref{subsec:hardware_design}. The hardware output will be gathered and regarded as the input of next block like FFN (feed forword network) in Transformer.

\section{Data Scheduler}
\label{subsec:data_scheduler}
Data scheduler is responsible for transforming the provided hybrid sparse attention mechanisms into the patterns that can be directly executed on the spatial accelerator of SALO, satisfying the dataflow constraint and the size constraint. We propose two techniques, namely data reordering and data splitting, to handle these two constraints respectively. We will first discuss our dataflow and analyze the data reuse given hybrid sparse attention patterns. Then we will introduce data reordering and splitting techniques to fully exploit the data reuse in hybrid sparse attention mechanisms.
\subsection{Dataflow and Data Reuse}
\label{subsubsection:dataflow_and_reuse}
We design a specific dataflow for SALO to maximize the data reuse when computing hybrid sparse attention mechanisms. This can help shorten the data path and reduce redundant memory access. To be more specific, SALO is equipped with highly optimized data path aiming at achieving a high data reuse rate among different queries within sliding window attention, and between sliding window attention and global attention. 

For data reuse within the sliding window attention, as discussed in Section \ref{subsec:sparse_attention_mechanisms}, most of the key vectors can be reused between successive queries. For example, if $q_i$ attends to the sequence of key vectors $[k_{i+a}, k_{i+a+1}, ..., k_{i+b-1}, k_{i+b}]$, $q_{i+1}$ should attend to $[k_{i+a+1}, k_{i+a+2}, \\..., k_{i+b}, k_{i+b+1}]$ because the window size $w=b-a+1$ is fixed. Thus there are $w-1$ reused key vectors. The data reuse of key vectors can be observed more intuitively through the visualized attention pattern. In Figure \ref{fig:star_transformer}, the sliding windows for $q_6$ and $q_7$ are highlighted in yellow and red respectively. And the corresponding key vectors are marked above. We can see that $k_6$ and $k_7$ highlighted in orange are shared between both windows, which can be reused.
To fully exploit data reuse in such sparse attention patterns, we propose a novel dataflow in which input $k$ and $v$ vectors are streamed into our spatial accelerator along the diagonal direction, and input $q$ vectors are streamed into our accelerator along the horizontal direction, as illustrated in Figure \ref{fig:PE}. With the proposed dataflow, the key vector that is no longer used ($k_{i+a}$ in the example) will be replaced by the new coming vector ($k_{i+b+1}$ in the example). The replacement of value vectors follows the same strategy as that of key vectors. 

For the data reuse between the sliding window attention and the global attention, the queries and keys of global tokens attend to all the keys and queries in the sequences based on the definition of global attention. At the same time, the sliding window attention always cover all the queries, keys and values in the sequence, which enables the global attention to be calculated without extra data access. This data reuse allows us to compute sliding window attention and global attention simultaneously with the same input vectors. The details about hardware implementation of dataflow will be discussed in Section \ref{subsubsec:dataflow_implementation}. 

\begin{figure}
\vspace{-0.3cm}
  \centering
    \includegraphics[width=0.43\textwidth]{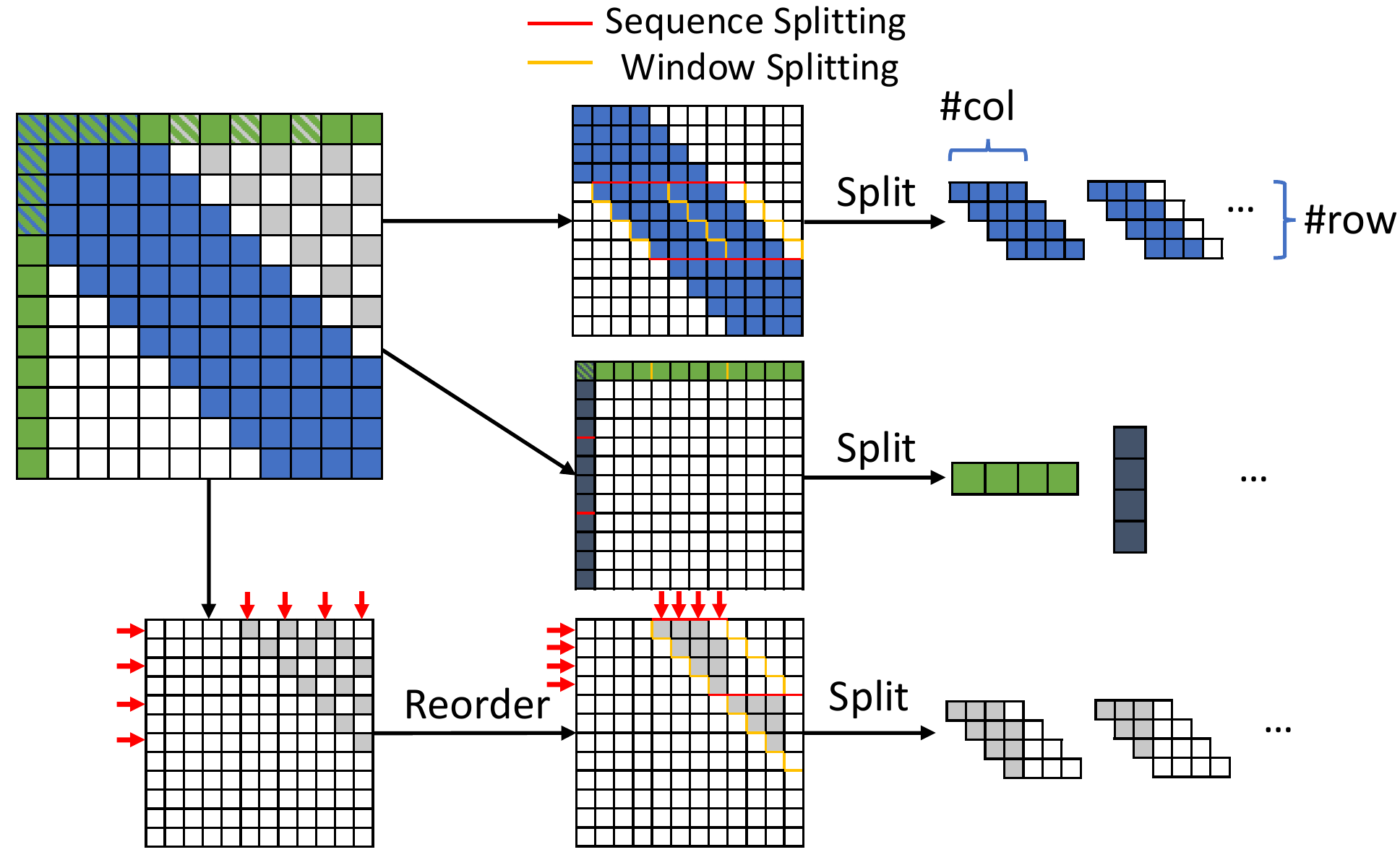}
    \caption{Data Reordering and Splitting in Data Scheduler}
    \label{fig:split_reorder}
    \vspace{-0.6cm}
\end{figure}
\subsection{Data Splitting and Reordering}
\label{subsubsection:data_splitting}
Based on our dataflow, we propose data splitting and reordering techniques to transform a specified hybrid sparse attention mechanism into patterns which satisfy our dataflow and can be directly processed by our accelerator. Figure \ref{fig:split_reorder} presents an overview for data splitting and reordering in the data scheduler. 

In practice, the sequence length $n$ and the window size $w$ are always too large for the accelerator to execute at once. Thus we propose \textit{data splitting} which divides input matrices into tiles and allows the attention to be computed in multiple passes. Data splitting consists of two parts, namely \textit{sequence splitting} and \textit{window splitting}. They split the pattern according to the size of PE array in the spatial accelerator. Both splitting strategies are also applied to the global attention because of their reuse dependency with the sliding window attention. 
Sequence splitting does not influence the calculation since the attention computations at different rows for different queries are independent. In contrast, window splitting divides computations of the same attention (for query $q_i$ at row $i$) into several parts. 
Then for query $q_i$, we obtain several output vectors (denoted as $output_i^1, output_i^2$) from split windows and need to accumulate these vectors for the final result. These output vectors stand for the results when the query attends to the keys from different parts (denoted as $T_1, T_2$) of the window. Recall the computation of the attention mechanism. For convenience, we denote the denominator in Eq. \ref{equ:softmax} as a weight $W$ ($W_1=\sum_{j\in T_1}\mathrm{exp}(S_{ij})$ and $W_2=\sum_{j\in T_2}\mathrm{exp}(S_{ij})$). Then we use a renormalizing transformation to recover the final result as if the window has not been split\footnote{The detailed derivation could be found in Appendix \ref{appendix:detailed_derivation}}:
\begin{equation}
\begin{aligned}
output_i 
=& \frac{\sum_{j\in T_1} \mathrm{exp}(S_{ij}) v_j + \sum_{j\in T_2} \mathrm{exp}(S_{ij}) v_j} {W_1+W_2} \\
= \frac{W_1}{W_1+W_2}\cdot & \frac{\sum_{j\in T_1} \mathrm{exp}(S_{ij}) v_j}{W_1} + \frac{W_2}{W_1+W_2}\cdot \frac{\sum_{j\in T_2} \mathrm{exp}(S_{ij}) v_j}{W_2} \\
=& \frac{W_1}{W_1+W_2}output_i^1 + \frac{W_2}{W_1+W_2}output_i^2
\end{aligned}    
\end{equation}

The dataflow in SALO only takes the data reuse for sliding window attention and global attention into consideration. Dilated window attention is not directly supported by SALO because of such a dataflow constraint. \textit{Data reordering} is a technique that transforms a dilated window attention to an equivalent sliding window attention, which is directly supported by the spatial accelerator of SALO, as shown in Figure \ref{fig:split_reorder}. After reordering, data reuse still exists for the dilated window attention, with a period $d$ (i.e., the dilation). For example, if $q_i$ attends to the sequence of key vectors $[k_{i+a}, k_{i+a+d}, ..., k_{i+b-d}, k_{i+b}]$, $q_{i+d}$ should attend to $[k_{i+a+d}, k_{i+a+2d},..., k_{i+b}, k_{i+b+d}]$, which is quite similar to the phenomenon in sliding window attention. So if we reorder Q matrix to group $q_i$ together with $q_{i+d}, q_{i+2d}, ...$, we can obtain a pattern, that can be executed on the spatial accelerator of SALO.

\section{Accelerator Design}
\label{subsec:hardware_design}
\subsection{PE Design}
As the main components of our spatial accelerators, the PE array, the global PE column and the global PE row share the same internal PE design. To minimize the cost of data transmission, our design enables the whole computation of attention to be finished within PEs: the matrix multiplication of $Q$ and $K$, softmax and the second matrix multiplication of $S'$ and $V$. 
\begin{figure}
\vspace{-0.4cm}
\includegraphics[width=0.46\textwidth]{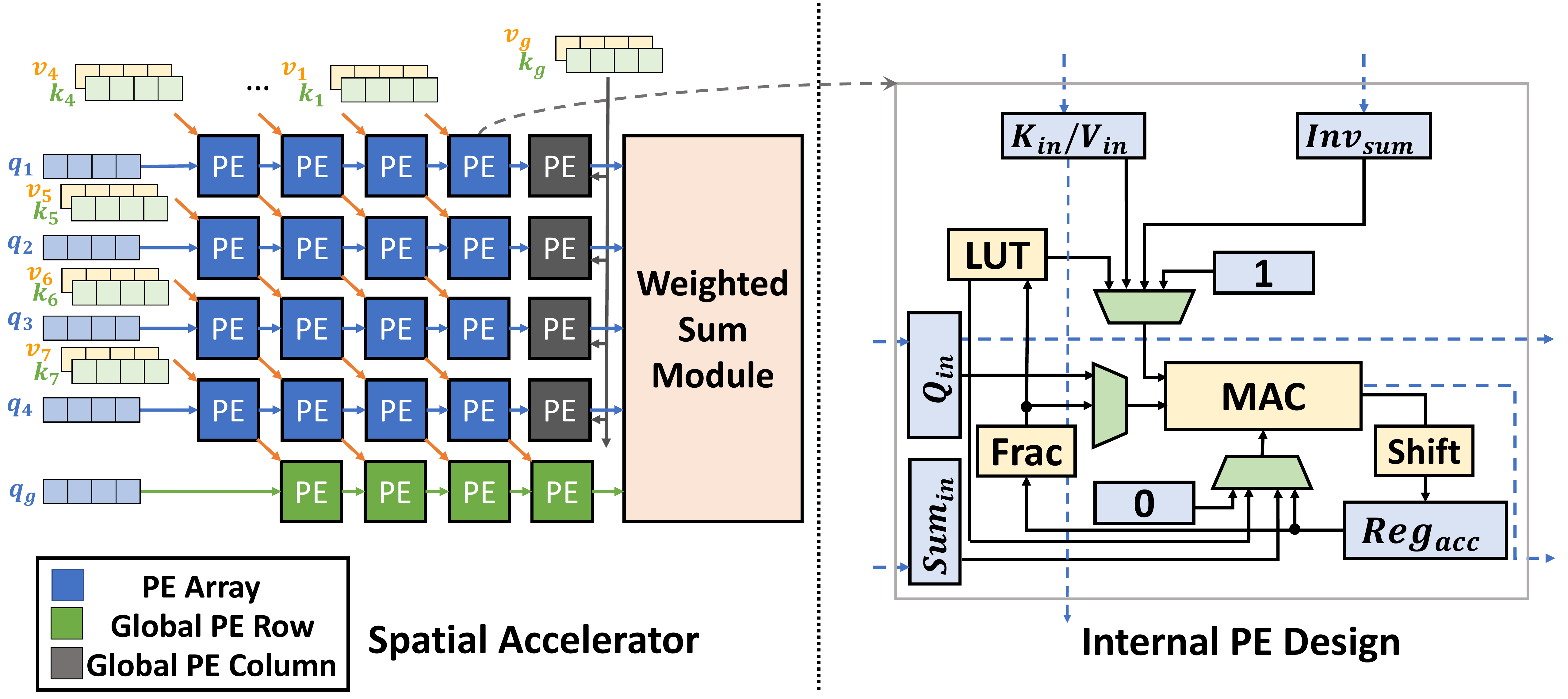}
\caption{The spatial accelerator and internal PE design} 
\label{fig:PE}
\vspace{-0.55cm}
\end{figure}

The right part of Figure \ref{fig:PE} shows the internal PE design. Each PE contains a fixed-point MAC as its computing unit, responsible for the calculation all through the computation of the attention mechanism. The PEs are grouped by rows. Each PE row accepts the query vector from the input port of the leftmost PE and emits the corresponding output vector to the rightmost PE.

\begin{figure}
\vspace{-0.3cm}
\includegraphics[width=0.43\textwidth]{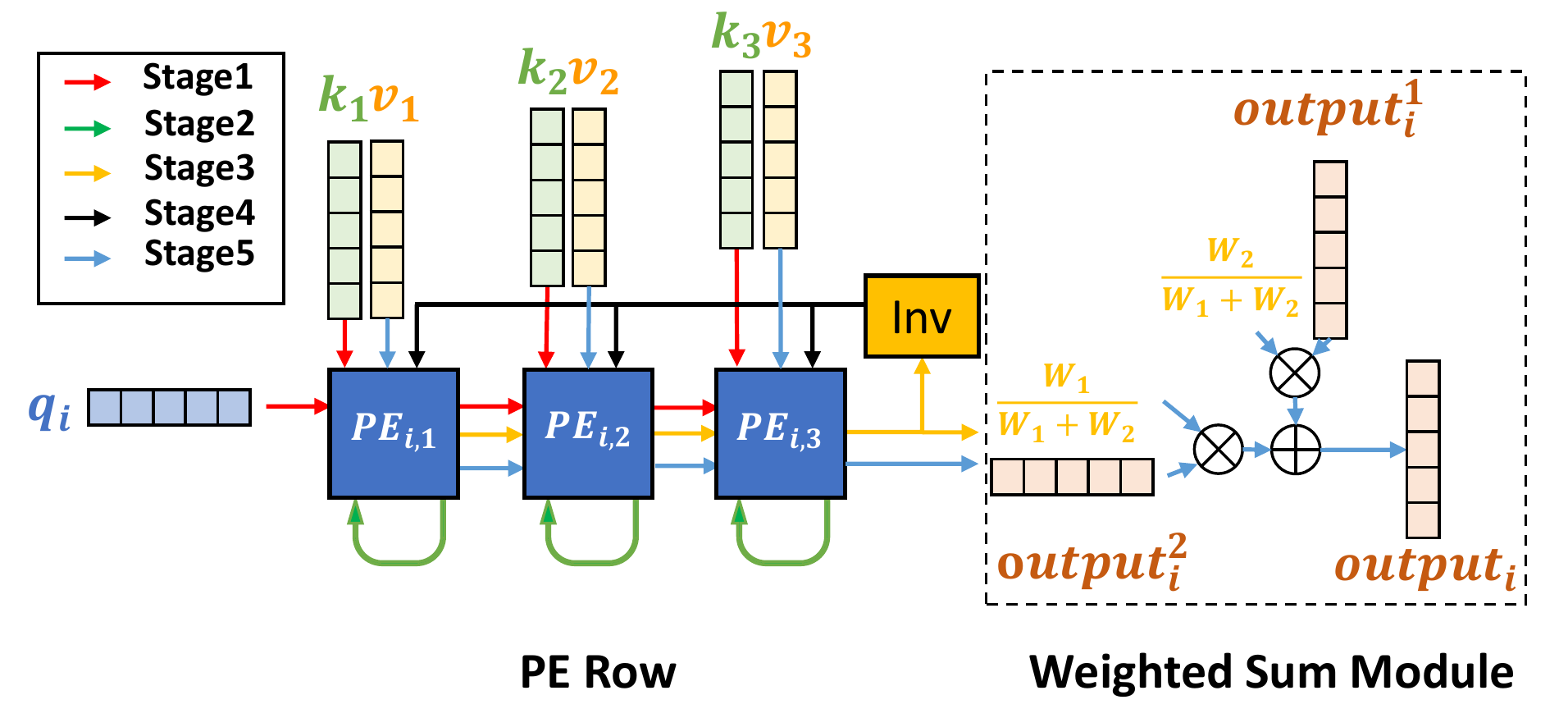}
\caption{The 5-stage data path} 
\label{fig:stage}
\vspace{-0.5cm}
\end{figure}

We divide the procedure of computation into 5 stages. The data paths at each stage are demonstrated in Figure \ref{fig:stage}. At stage 1, the matrix multiplication of $Q$ and $K$ is done in a typical output stationary systolic manner. $\mathrm{PE}_{i,j}$ receives an element of $q_i$ and an element of $k_j$ each cycle. Then the PE multiplies them and accumulates the result to $Reg_{acc}$. The received element will be passed to a neighboring PE next cycle. Stage 1 continues until the computation is completed, when $\mathrm{PE}_{i,j}$ holds the element $S_{ij}$ in $Reg_{acc}$. At stage 2, the exponential of $Reg_{acc}$, i.e., $\mathrm{exp}(S_{ij})$ will be calculated. Here we follow the method proposed in Softermax\cite{stevens2021softermax}, which utilizes a piece-wise linear function to approximate the exponential function because the linear function can be calculated efficiently using the MAC unit. Two lookup tables (LUT in Figure \ref{fig:PE}) are used to store the slope and the y-interception for each segment. The result will be stored back to $Reg_{acc}$. At stage 3, the result in the PEs from the same row will be accumulated to get $\sum_k \mathrm{exp}(S_{ij})$, which is the denominator of softmax. The accumulation can be finished with a horizontal dataflow: each PE receives the partial sum from PE on the left, updates the sum with the exponential stored in $Reg_{acc}$ and passes the sum to the PE on the right. When the sum leaves the rightmost PE, the inverse of the sum, $(\sum_k \mathrm{exp}(S_{ij}))^{-1}$ will be calculated and broadcast back to PEs in the row. We choose such a design because the circuits of divider is complex, causing significant cycle time and area costs. At stage 4, each PE multiplies the $\mathrm{exp}(S_{ij})$ stored in $Reg_{acc}$ and the received inverse to obtain $S'_{ij}$. Then at stage 5, the value vector $v_j$ is fed into $\mathrm{PE}_{i,j}$ from the same port as the key vector. Different from the output stationary dataflow in stage 1, the PE row applies a weight stationary dataflow: $\mathrm{PE}_{i,j}$ receives a partial sum from the PE on the left, updates it with the product of element of $v_j$ and $S'_{ij}$, and then passes it to the PE on the right. The output of the PE row will enter the weighted sum module to be merged with the former output.

\subsection{Dataflow Implementation}
\label{subsubsec:dataflow_implementation}

To implement the dataflow proposed in Section \ref{subsubsection:dataflow_and_reuse}, we utilize the diagonal connections between PEs in the PE array for the data stream of key vectors and value vectors, as is shown in Figure \ref{fig:PE}. When the key vectors or value vectors transfer to a new PE row along the diagonal direction, the vector in the rightmost PE is dropped, and a new vector enters the PE row from the leftmost PE. Thus we could realize the data reuse between successive queries of sliding window attention. As for the data reuse of global attention, we attach a global PE row and a global PE column to the PE array. The global PE row is responsible for calculating the attention of the queries of global tokens. So the global PE row reuses the key vectors and value vectors from the PE array. Similarly, the global PE column reuses the queries from the PE array. One may ask how to support multiple global tokens with a single PE row or column. In practice, the number of the global tokens $n_g$ is far less than the sequence length $n$ and the window size $w$.
But at the same time, because of the data splitting, each input vector(query, key, value) is loaded to the array more than once. In conclusion, we can support up to $n_g =min\{\lceil n / \#row\rceil, \lceil w / \#col\rceil\} $ global tokens with a single PE row and column, which is sufficient in real workload.

\subsection{Weighted Sum Module}
The weighted sum module works as a postprocessing module for the output from the PEs to implement the algorithm described in Section \ref{subsubsection:data_splitting}. At the third stage shown in Figure \ref{fig:stage}, the sum of exponential $W_1$ is collected. With the weight $W_1$ and the weight of previous output $W_2$, we can calculate the corresponding normalized weights $\frac{W_1}{W_1+W_2}$ and $\frac{W_2}{W_1+W_2}$. Then in each cycle of stage 5, an element of the output vector will arrive at the weighted sum module. The weighted sum of the elements from the two output vectors will be computed to obtain an output vector with weight $W_1+W_2$. At the hardware level, this procedure needs two multipliers, and an adder for each PE row, which is an acceptable cost.

\section{Evaluation}
\subsection{Experiment Settings}
\textbf{Models and Datasets.} 
To evaluate the performance of SALO and its speedup ratio compared to GPU and CPU, we use two representative models, Longformer \cite{beltagy2020longformer} and ViL \cite{zhang2021multi} as the benchmark. Longformer is a variant of Transformer designed for long document in natural language processing. We choose the most commonly used implentation from Transformers \cite{wolf2020transformers} by Huggingface. ViL is a variant of ViT \cite{dosovitskiy2020image} extracting multi-scale features from images with sparse hybrid mechanisms. We use the implementation open sourced by the author. As for the hyper parameters, we follow the settings of Longformer-Base-4096 and ViL-Medium-Wide. Note that ViL consists of 4 stages with an attention layer in each stage.  But we only test the performance on the first 2 stages that involve hybrid attention mechanism. We use IMDB and Hyperpartisan to evaluate the accuracy of Longformer and ImageNet-1K to evaluate the accuracy of ViL.

\noindent\textbf{Platform.}
To obtain the speedup of SALO, we choose CPU and GPU as the baseline. We test the model performance and the corresponding power consumption on a server CPU (Intel Xeon E5-2630 v3) and a server GPU (GTX 1080Ti). All the experiments of CPU and GPU are conducted with PyTorch \cite{paszke2019pytorch} 1.5.0. And we use MKL for CPU and CUDNN for GPU as the backend, which are specially optimized for the architecture.

\noindent\textbf{Hardware Synthesis}
We implement the spatial accelerator of SALO in Chisel \cite{bachrach2012chisel}. The design in Chisel is further compiled to Verilog and synthesized using Synopsys DC compiler with FreePDK 45nm technology to obtain the report of power and area. Table \ref{tab:synthesis} details the parameters and result of the synthesis. For performance analysis, we extend the cycle-accurate performance model from Sanger\cite{lu2021sanger} while taking different hybrid sparse attention mechanisms, data splitting and reordering involved in the data scheduler into consideration. 

\begin{table}[t]
\vspace{-0.2cm}
\caption{Synthesis Details}
\label{tab:synthesis}
\begin{tabularx}{\linewidth}{YY}
\toprule
\multicolumn{2}{c}{\textbf{Hardware Paramters}}                 \\ 
\midrule
\multicolumn{1}{c}{PE array size}       & $32\times 32$       \\ 
\multicolumn{1}{c}{Global PE column}    & 1                   \\ 
\multicolumn{1}{c}{Global PE row}       & 1                   \\ 
\multicolumn{1}{c}{Weighted Sum Module} & 33                  \\
\multicolumn{1}{c}{Query Buffer} & 16KB \\
\multicolumn{1}{c}{Key Buffer} & 32KB \\
\multicolumn{1}{c}{Value Buffer} & 32KB \\
\multicolumn{1}{c}{Output Buffer} & 32KB \\
\toprule
\multicolumn{2}{c}{\textbf{Synthesis Report}}             \\
\midrule
\multicolumn{1}{c}{Frequency}           & 1GHz \\
\multicolumn{1}{c}{Power}               & 532.66mW            \\ 
\multicolumn{1}{c}{Area}                & 4.56$\mathrm{mm}^2$ \\ 
\bottomrule
\end{tabularx}
\vspace{-0.5cm}
\end{table}

\begin{table}[b]
\vspace{-0.3cm}
\caption{Key parameters of attention layers}
\label{tab:param}
\begin{tabularx}{\linewidth}{YYYYYY}
\toprule
Parameters   & Sequence length & Window size & Hidden size  & Global Token & Sparsity\\
\midrule
\small Longformer & 4096 & 512 & 768 & 1 & 0.125\\
\scriptsize  ViL-stage1 & $56\times56$ & $15\times 15$ & 192 & 1 & 0.072\\
\scriptsize  ViL-stage2 & $28\times28$ & $15\times 15$ & 384 & 1 & 0.288\\
\bottomrule
\end{tabularx}
\end{table}

\subsection{Performance Analysis}
For both models, we test the latency and the power consumption among CPU, GPU and SALO under the same workload. Because the target of SALO is to accelerate the hybrid attention mechanism, we split the single attention layer from the original models as the workload. We use one attention layer from Longformer, and two attention layers from the first two stages of ViL. The key parameters of all three kinds of attention layers are summarized in Table \ref{tab:param}.

\begin{figure}[t]
\begin{subfigure}{0.23\textwidth}
\includegraphics[width=\textwidth]{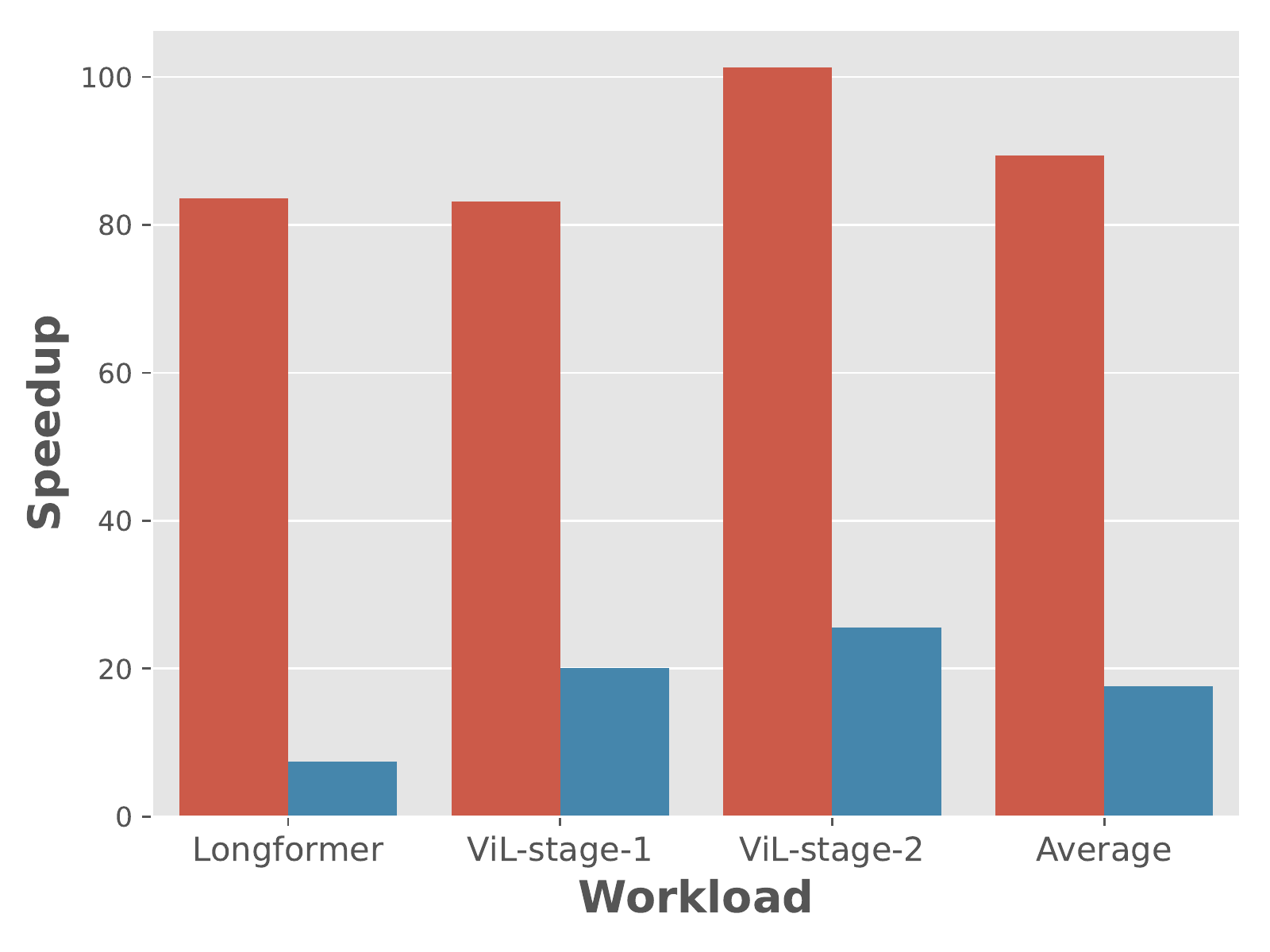}
\caption{} 
\label{fig:speedup}
\end{subfigure}
\begin{subfigure}{0.23\textwidth}
\includegraphics[width=\textwidth]{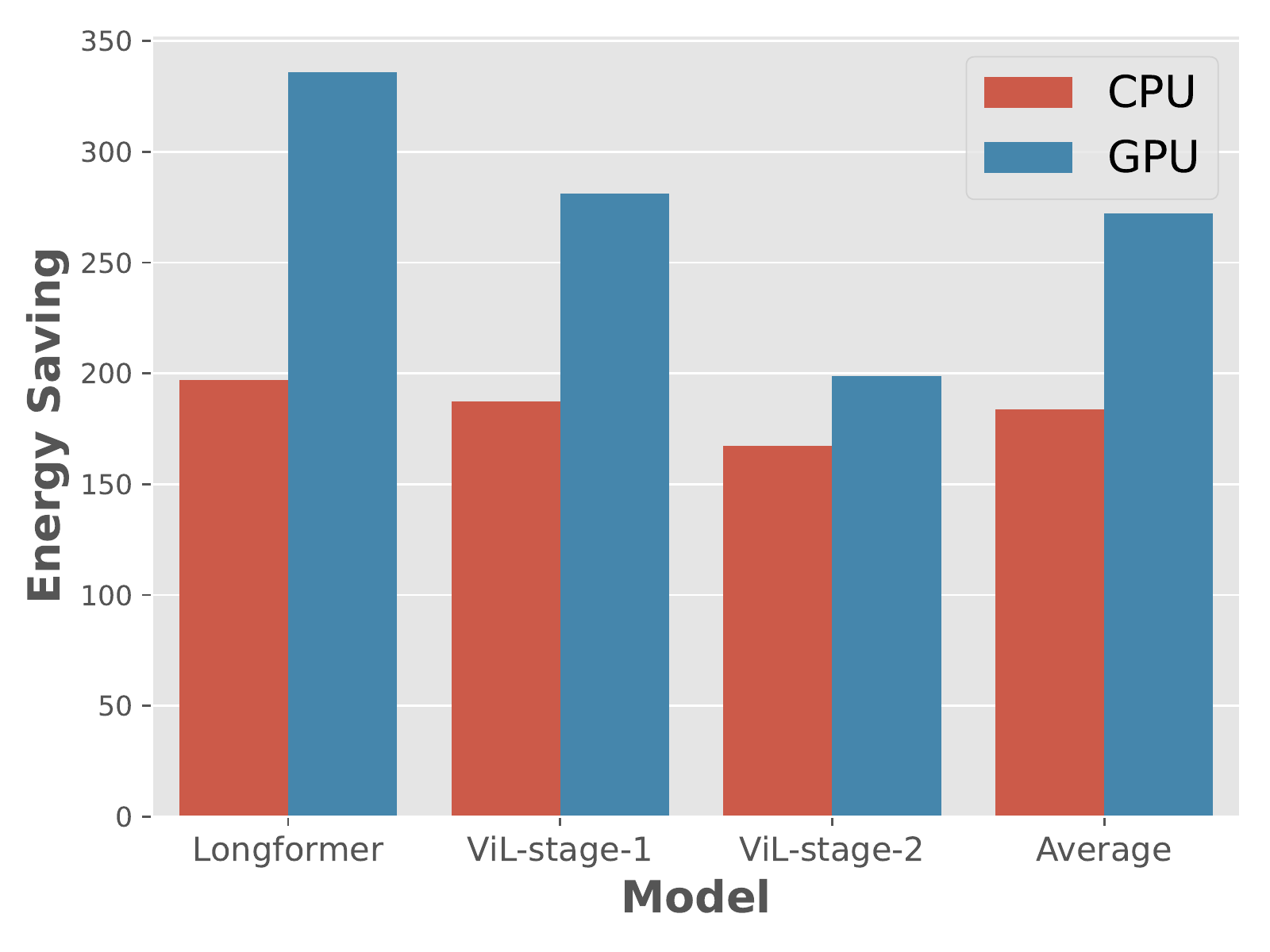}
\caption{} 
\label{fig:power}
\end{subfigure}
\caption{The speedup and energy saving of SALO compare to CPU and GPU}
\vspace{-0.3cm}
\end{figure}

As shown in Figure \ref{fig:speedup}, SALO achieves $83.57\times$, $83.12\times$, $101.31\times$ speedup compared to CPU ($89.33\times$ on average), and $7.38\times$, $20.10\times$, $25.51\times$ speedup compared to GPU ($17.66\times$ on average). The speedup comes from our specially designed dataflow for hybrid sparse attention mechanisms which maximizes the data reuse in PE array. As for CPU and GPU, the sliding window attention cannot be directly supported by the highly optimized GEMM implementation. 

Energy is another important metric to evaluate the hardware efficiency. From Figure \ref{fig:power}, we could see that SALO achieves $196.90\times$, $187.53\times$, $167.15\times$ energy saving compared to CPU ($183.86\times$ on average), and $336.05\times$, $281.29\times$, $198.78\times$ energy saving compared to GPU ($272.04\times$ on average). The extremely low power design makes it possible to integrate SALO into existing architectures to perform hybrid attention mechanisms efficiently.

\vspace{-0.2cm}
\subsection{Comparison with Existing Accelerator}

Sanger\cite{lu2021sanger} is the state-of-the-art accelerator design that aims at accelerating dynamic sparse attention mechanisms. Sanger also uses a systolic array as the main part of its design, and has a similar number of PEs ($64\times 16$) with SALO. The peak performance of Sanger and SALO are nearly equal. However, the performance of Sanger is limited in the case of long input sequences. This is because Sanger has an extra step before using the systolic array to calculate sparse attention, that is to predict the score matrix $\hat{S}$ with low-precision. This latency occupies a quadratic complexity, regardless of the sparsity. Moreover, the utilization ratio of Sanger's PE array (around $55\%$ to $75\%$) is smaller than ours (>75\%) given the same sparsity ratios (around 0.05-0.30). This is because Sanger applies irregular sparse patterns while the PEs of our accelerator is almost fully utilized given hybrid sparse patterns. 
With the same number of PEs, the same sparsity and frequency, SALO outperforms Sanger with $1.33\times$ speedup for its higher utilization and shorter data path without a predicting step.

\subsection{Impact of Quantization}
Because SALO does not introduce any modification to attention mechanisms, there will not be an approximation error when using SALO to accelerate the inference. But SALO uses a low precision fixed point arithmetic to achieve higher performance that quantizes the Query, Key, Value matrix to 8 bits (4 bits for fraction part). As a result, the output of SALO is in 16 bits, which may potentially influence the accuracy. Thus we conduct experiments on Longformer and ViL to measure the impact of quantization. We slightly modified the self attention layers in Longformer and ViL by inserting quantization layers after the operators, to simulate the precision on SALO. These quantization layers are implemented by QPyTorch\cite{zhang2019qpytorch}, a low-precision arithmetic simulation package. We perform quantization-aware finetuning on both pretrained models.

\begin{table}[t]
\caption{Comparison between the original model and quantized model}
\label{tab:quant}
\begin{tabularx}{\linewidth}{YYYY}
\toprule
Model     & \multicolumn{2}{c}{Longformer}            & ViL         \\
\midrule
Datasets  & \multicolumn{1}{c}{IMDB}  & Hyperpartisan & ImageNet-1K \\ 
\midrule
Original  & \multicolumn{1}{c}{95.34} & 93.42         & 82.87       \\
Quantized & \multicolumn{1}{c}{95.20} & 93.46         & 82.80       \\
\bottomrule
\end{tabularx}
\end{table}

Through these experimental results shown in Table \ref{tab:quant}, we demonstrate that the quantization in SALO will not cause significant degradation on accuracy in both models, which enables users to deploy the pretrained model to SALO without much effort.

\section{Conclusion}
In this paper, we propose SALO, an efficient spatial accelerator that enables hybrid sparse attention mechanisms for long sequences. SALO contains a data scheduler which transforms hybrid sparse attention mechanisms and a spatial accelerator which perform sparse attention computations efficiently. Experiments show that SALO can achieve high performance and low energy consumption without accuracy loss.

\begin{acks}
This work is sponsored by the National Natural Science Foundation of China (NSFC) (62102249, 61832006) and Shanghai Pujiang Program (21PJ1408200).
\end{acks}

\bibliographystyle{ACM-Reference-Format}
\bibliography{ref}

\appendix
\section{Detailed derivation}
\label{appendix:detailed_derivation}
In Section \ref{subsubsection:data_splitting}, we have introduced the transformation to recover the final result based on the results of different windows. Due to the page limit, we omit some of the derivation. Here we will give the detailed derivation of the equation.

First, we denote the denominator in Eq. \ref{equ:softmax} as weight $W$. We denote each part in window splitting as a set of tokens, namely $T_k$. Similarly, we can define the weight of each part as $W_k$.
\begin{equation}
\begin{aligned}
W_k=\sum_{j\in T_k}\mathrm{exp}(S_{ij})
\end{aligned}    
\end{equation}

Clearly, the weight $W$ of the whole window is the sum of the weight $W_k$ of each part.

\begin{equation}
\begin{aligned}
W&=\sum_{j\in T_1\cup ... \cup T_K}\mathrm{exp}(S_{ij})\\
&=\sum_{j\in T_1}\mathrm{exp}(S_{ij}) + ... + \sum_{j\in T_K}\mathrm{exp}(S_{ij})\\
&= \sum_k W_k
\end{aligned}    
\end{equation}

According to the expression of attention mechanism, for each window, the output vector can be calculated as following.

\begin{equation}
\begin{aligned}
output_i = \frac{\sum{\mathrm{exp}(S_{ij}) v_j}}{\sum {\mathrm{exp}(S_{ij})}}
\end{aligned}    
\end{equation}

Now we can build a bridge between the result of each part $output_i^k$ and the final result $output_i$.
\begin{equation}
\begin{aligned}
output_i =& \frac{\sum{\mathrm{exp}(S_{ij}) v_j}}{\sum {\mathrm{exp}(S_{ij})}} \\
=& \frac{\sum_{j\in T_1} \mathrm{exp}(S_{ij}) v_j + ... + \sum_{j\in T_K} \mathrm{exp}(S_{ij}) v_j}{W} \\
=& \frac{\sum_{k}(\sum_{j\in T_k} \mathrm{exp}(S_{ij}) v_j)}{W} \\
=& \sum_k \frac{\sum_{j\in T_k} \mathrm{exp}(S_{ij}) v_j}{W}\\
=& \sum_k \frac{W_k}{W}\cdot \frac{\sum_{j\in T_k} \mathrm{exp}(S_{ij}) v_j}{W_k}\\
=& \sum_k \frac{W_k}{W}\cdot output_i^k
\end{aligned}    
\end{equation}

So, when applying window splitting to a given sparse attention pattern, we only need to store the weight and the output vector of each part to recover the final result.
\end{document}